\documentclass{emulateapj}
\usepackage{longtable}
\usepackage{subfigure}
\usepackage{amsmath}
\usepackage{mathtools}

\providecommand{\e}[1]{\ensuremath{\times 10^{#1}}}

\begin{document}

\title{The Case for a Low Mass Black Hole in the Low Mass X-ray Binary V1408 Aquilae (= 4U 1957+115)}

\author{Sebastian Gomez\altaffilmark{1}, 
             Paul A. Mason\altaffilmark{1,2},
             and Edward L. Robinson\altaffilmark{3}}

\altaffiltext{1}{Department of Physics, 
                 University of Texas at El Paso, 
                 El Paso, TX 79968, USA}
\altaffiltext{2}{Department of Mathematics and Physical Science, DACC, 
                 New Mexico State University, 
                 Las Cruces, NM 88003, USA}
\altaffiltext{3}{Department of Astronomy, 
                 University of Texas at Austin, 
                 1 University Station,
                 Austin, TX, 78712, USA}

\begin{abstract}

There are very few confirmed black holes with a mass that could be $\sim\! 4\, M_\odot$ and no neutron stars with masses greater than $\sim\! 2\, M_\odot$, creating a gap in the observed distribution of compact star masses. Some black holes with masses between 2 and $4\, M_\odot$ might be hiding among other X-ray sources, whose masses are difficult to measure. We present new high-speed optical photometry of the low-mass X-ray binary V1408 Aql (= 4U 1957+115), which is a persistent X-ray source thought to contain a black hole. The optical light curve of V1408~Aql shows a nearly sinusoidal modulation  at the orbital period of the system superimposed on large  night-to-night variations in mean intensity. We combined the new photometry with previously-published photometry to derive a more precise orbital period, $P = 0.388893(3)$\ d, and to better define the orbital light curve and night-to-night variations. The orbital light curve agrees well with a model in which the modulation is caused entirely by the changing aspect of the heated face of the secondary star. The lack of eclipses rules out orbital inclinations greater than $65^{\circ}$. Our best models for the orbital light curve favor inclinations near $13^{\circ}$ and black hole masses near $3\, M_\odot$ with a 90\% upper bound of $6.2\, M_\odot$, and a lower bound of $2.0\, M_\odot$ imposed solely by the maximum mass of neutron stars. We favor a black hole primary over a neutron star primary based on evidence from the X-ray spectra, the high spin of the compact object, and the fact that a type I X-ray burst has not been observed for this system. Although uncertainties in the data and the models allow higher masses, possibly much higher masses, the compact star in V1408~Aql is a viable candidate for a black hole lying in the mass gap.

\end{abstract}

\keywords{binary systems: X-Ray binaries, individual (\objectname{4U 1957+115})}

\maketitle 

\section{Introduction}
Low mass X-ray binaries (LMXB) are comprised of a compact stellar remnant, either a black hole or a neutron star, that accretes matter from a low-mass Roche-Lobe-filling secondary star via a circumstellar disk. The LMXB 4U 1957+115 was first detected by the Uhuru satellite \citep{gia74} and its optical counterpart V1408 Aql was first identified by \citet{mar78}. The X-ray properties of 4U 1957+115 are unusual. LMXBs usually cycle between active and quiescent states, but 4U 1957+115 has been persistently active for more than 40 years, the longest interval of any known LMXB, remaining in a spectrally soft, disk-dominated X-ray state with no detectable radio jet \citep{wij02,rus11}. V1408 Aquilae may be the only BH LMXB, with the possible exception of LMC X-3, that has been found persistently active. We note that LMC X-1, LMC X-3, and Cyg X-1 \citep{rus10} are persistent high mass X-ray binary systems.

The X-ray light curve of the system has not shown any orbital modulation \citep{wij02}, but optical observations by \cite{tho87} revealed a nearly sinusoidal orbital variation with a peak-to-peak amplitude of 23\% and a period of 9.33 hr. \cite{hak99} observed the light curve on two nights and saw a change in the shape of the light curve and a significant increase in the amplitude of the variations. Further evidence of night-to-night variations was presented by \cite{rus10} from 144 images of the source spread over three years. The larger data sets of \cite{bay11} and \cite{mas12} confirmed both the sinusoidal orbital modulation and the changes in mean brightness from night to night. Following \citet{tho87}, \citet{bay11} showed that the orbital light curve can be reproduced by a model in which the secondary star is heated by flux from the accretion disk. The orbital modulation is produced entirely by the heated face of the secondary star as it rotates into and out of view. Our results differ from the models by \cite{hak14} in that their models do not find any evidence for a secondary star in the spectral energy distribution, and do not agree with a secondary whose face is irradiated by X-rays. The SED measured in this work is consistent with our model. Since the heated face of the secondary would be just a small, high-temperature black-body	perturbation on the SED and would not be distinguishable in the measured SED.

Several lines of evidence indicate that 4U 1957+115 has a black hole primary. It is common for LXMBs with neutron star primaries to show type I X-ray bursts but 4U 1957+115 has never shown such bursts. The X-ray spectrum is well described by a multi-temperature blackbody, with an additional non-thermal power law component in 15\% of the observations \citep{now12}. Its high inner disk temperature and small inner disk radius are consistent with a black hole primary \citep{whi83,wij02,now08,rus10,now12}. Models of the X-ray spectra that allow for rotation of the primary star yield large spin rates: from $a^* \gtrsim 0.9$ for a black hole with a mass of $3\, M_\odot$ and a distance of 10 Kpc, to a near-maximal spin of $a^* \approx 1$ for larger values of the mass and distance \citep{now12}. The observed spins of black holes range up to $a^*\approx1$, while the spins of neutron stars are exclusively less than $a^*= 0.1$  \citep{mil11}. \cite{yaq93}, \cite{sin94}, and \cite{bay11} argued that the primary in 4U 1957+115 is a neutron star, in part because the mass ratio $q = M_2/M_1$ is large, suggesting a small-mass primary that is more consistent with a neutron star than a black hole. We note, however, that a black hole with an unusually low mass could also yield a high mass ratio. 

The distribution of the known black hole and neutron star masses has a gap between $2\, M_\odot$ and $\sim4\, M_\odot$. Among the neutron stars with reliably measured masses, the most massive is J0348+0432 at $2.01\pm0.21, M_\odot$ followed by J1614+2230 at $1.97\pm0.04\, M_\odot$ \citep{dem10,ant13,lat14}. The least massive black hole, GRO J0422+32 has a mass of $3.97\pm0.95\, M_\odot$ \citep{gel03}, followed by GRS 1009-45 with a mass of near $4.4\, M_\odot$ and not less than $3.64\, M_\odot$ \citep{fil99}, and possibly by 4U~1547-47 with an estimated mass of $\lesssim 4\, M_\odot$  \citep{kre12}. A very massive neutron star or a very low mass black hole could be produced from a progenitor with a mass of $\sim 22M_\odot$ \citep{fry99}. The probability that the observed gap between the masses of neutron stars and black holes is a mere statistical fluke is low \citep{oze10,far11}, although it remains possible that the gap has a non-zero but sparse population.

If the mass gap is real, it has important implications for the physics of core-collapse supernovae \citep{fry12}. \cite{bel12} proposed a theoretical explanation for the existence of the mass gap that depends on the growth time of instabilities that lead to core-collapse supernovae. Stars in the $20 - 40\, M_\odot $ are the ones likely to have high mass neutron stars or low mass black holes as remnants that would lie in the mass gap. If the growth time of instabilities is larger than 200 milliseconds, a continuous distribution would be expected in this range. But models that assume a growth time of 10 - 20 milliseconds introduce and explain the observed mass gap \citep{bel12}. The observed mass distribution might, however, be subject to strong selection effects. One possibility is that low-mass black holes are hiding among other X-ray sources whose masses are notoriously difficult to measure. The absorption-line spectra of their secondary stars are generally not visible. \citet{oze10} concluded that there are simply not enough persistent systems for this to be the sole source of the gap: Even if every X-ray binary system that could contain a black hole were to contain a low-mass black hole, the gap would still not be fully populated. Nevertheless, it remains possible that a few of the missing low-mass black holes are lurking among transient X-ray sources.

In this paper we present new optical photometry of V1408 Aql. We derive an improved orbital ephemeris for the system and model the mean orbital light curve using our \texttt{XRBinary} light curve synthesis code, from which we constrain the orbital inclination and the mass of the compact star. We find that the most likely mass for the compact star places it inside the gap in the mass distribution, although the range of possible masses is large. In section 2 we describe the observations and summarize the behavior of the light curve, and in section 3 we derive an updated orbital ephemeris. In section 4 we discuss the models of the optical light curve and in section 5 we discuss constraints on the distance to the source.

\begin{table}[ht]
	\caption{Journal of Observations} 
	\begin{center}
		\begin{tabular}{cccccc}
			\hline\hline
			UT Date 		& UTC Start & Duration (hr)	 \\
			\hline
			14-July-2012	&  	07:06  	&  3.6			 \\
			15-July-2012	&  	06:25  	&  4.5			 \\
			16-July-2012	&	06:31  	&  4.2			 \\
			17-July-2012	& 	07:39  	&  3.1			 \\
			18-July-2012	&  	06:46  	&  4.0			 \\
			11-August-2012	&  	02:41  	&  4.7			 \\
			12-August-2012	&  	02:29  	&  4.5			 \\
			13-August-2012	&  	02:30  	&  4.0			 \\
			15-August-2012	&  	02:51  	&  4.9			 \\
			\hline
		\end{tabular}
	\end{center}
	
	\raggedright
	$^{a}$Time resolution of all data is 10s.
	\label{journal-tab} 
\end{table} 

\begin{figure}
	\begin{center}
		\includegraphics[angle=0.0,scale=0.27]{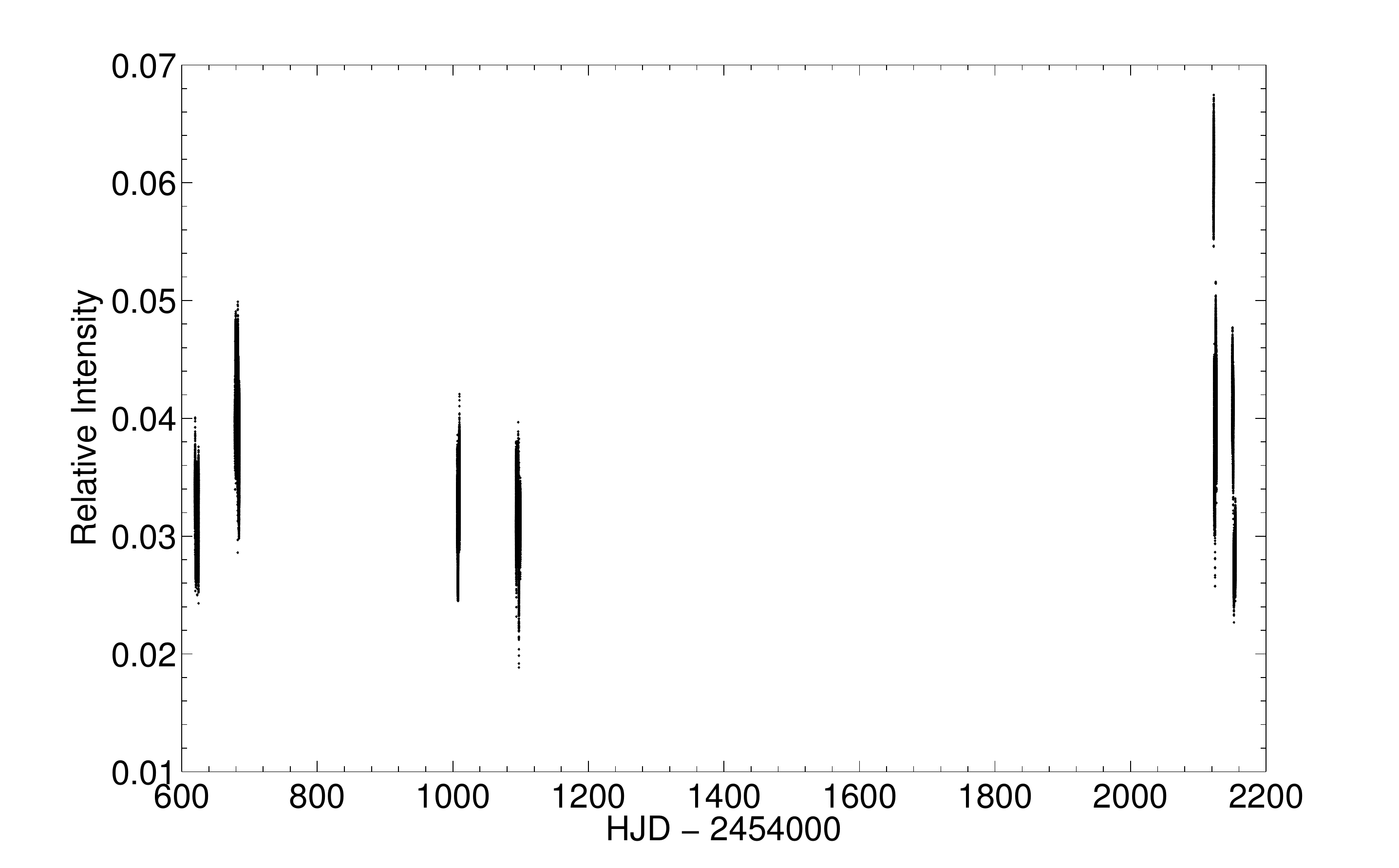}
	\end{center}
	\caption[New 2008 light curve.]{Twenty nine nights of photometry of V1408 Aql. On the scale of this figure the individual nights are not resolved, only entire observing runs. The data in the four clumps on the left side of the figure were obtained by \cite{bay11} and \cite{mas12}, while the new data comprise the two clumps on the right side of the figure. The detached clump of bright points on the right side of the figure come from a single night, HJD~2456122}.
	\label{alldata-fig}
\end{figure}

\begin{figure}
	\centering
	\subfigure[July 2012 lightcurves]
	{
		\includegraphics[scale=0.27]{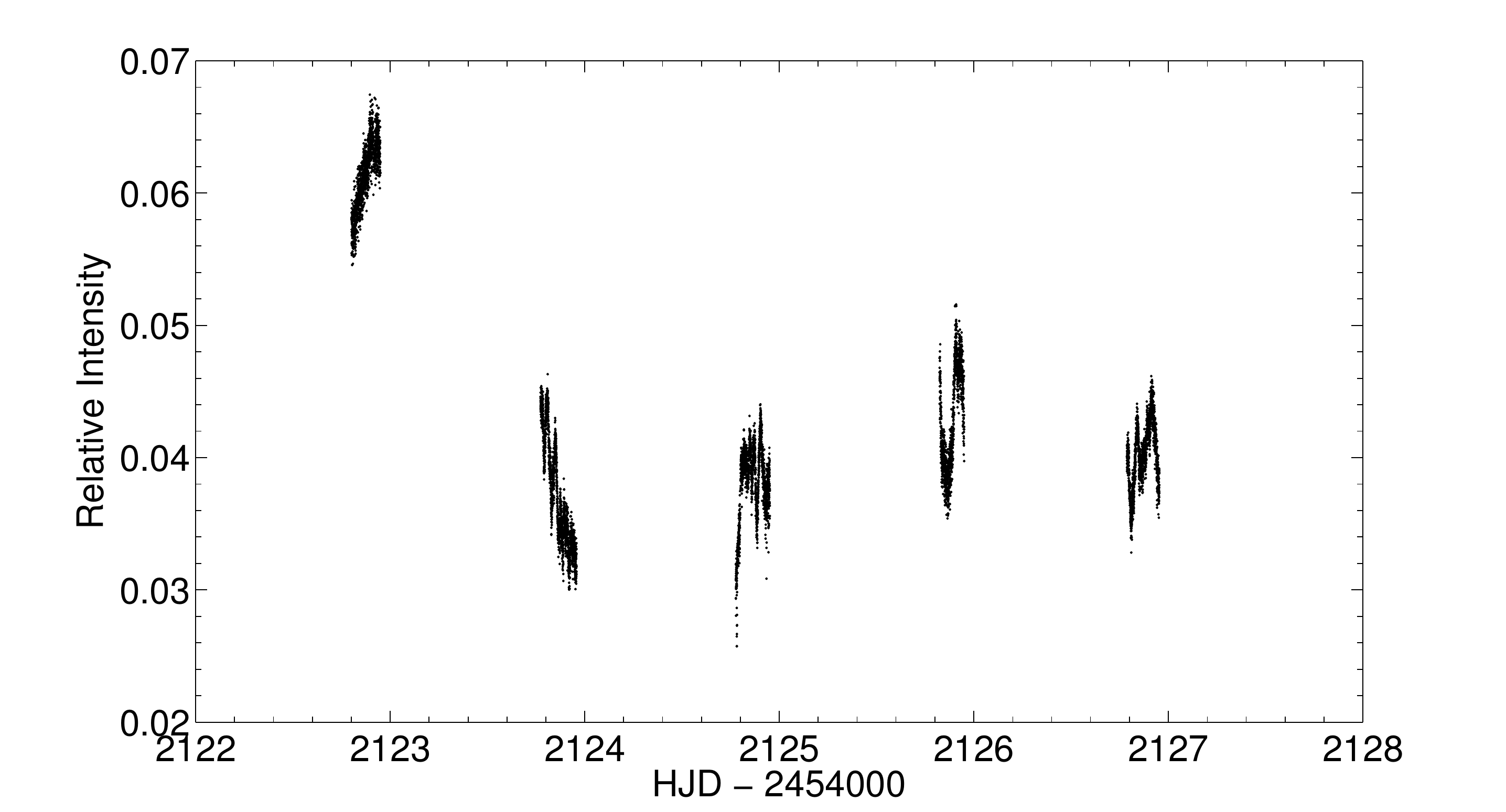}
		\label{onea}
	}
	\\
	\subfigure[August 2012 lightcurves]
	{
		\includegraphics[scale=0.27]{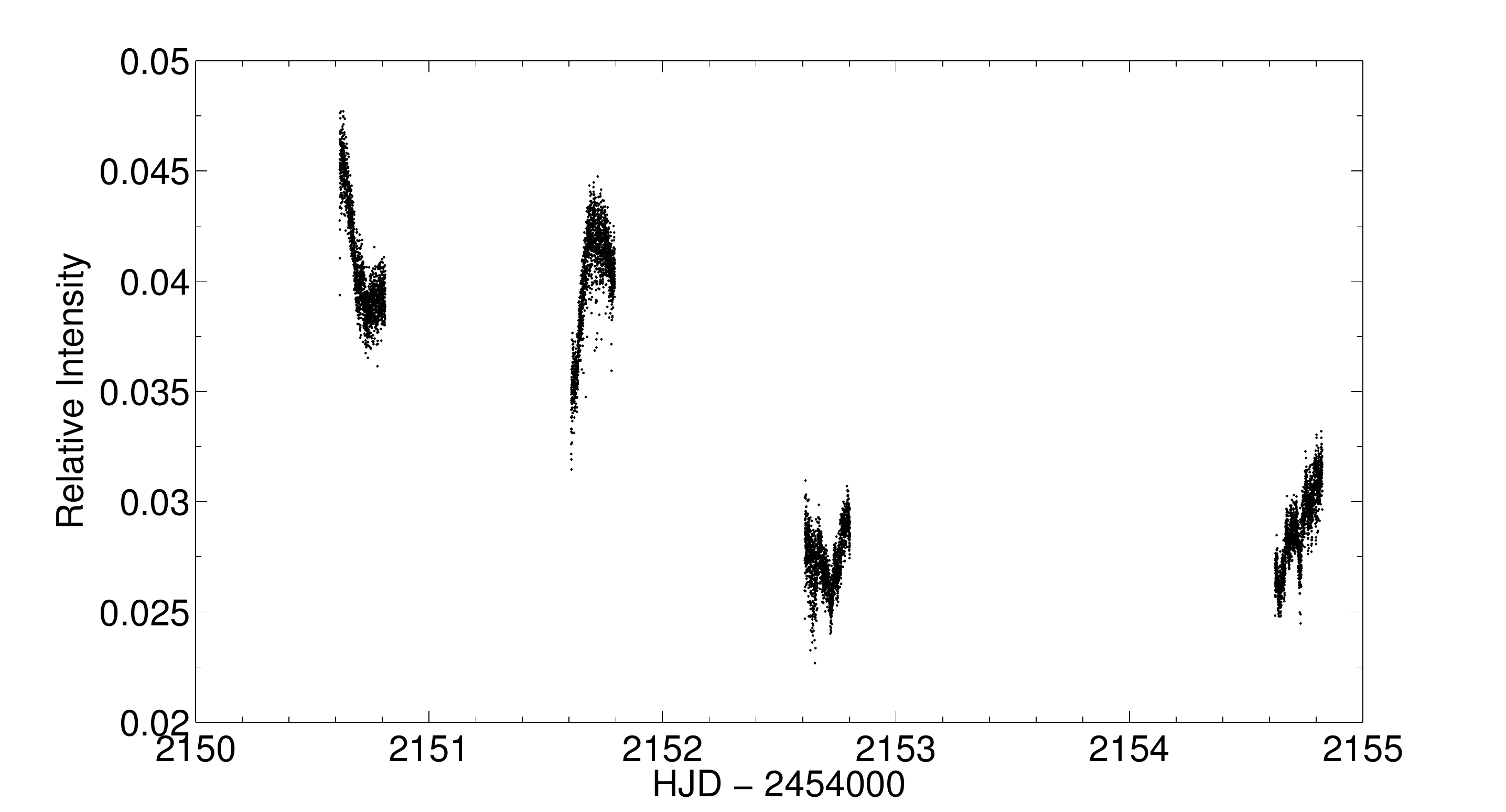}
		\label{oneb}
	}
	\caption{The light curves of V1408 Aql on five nights in July 2012 (top) and four nights in August 2012 (bottom). }
	\label{runs-fig}
\end{figure}

\section{The Optical Orbital Light Curve}

We obtained new high-speed optical photometry of V1408 Aql with the Argos CCD photometer on the 2.1m Otto Struve telescope at McDonald Observatory \citep{nat04}. The photometer produces a sequence of consecutive CCD images, all with 10-second exposure times for the data reported here. We observed V1408 Aql on five nights in July 2012 and four nights in August 2012 for 3 - 5 hours per night (see Table~\ref{journal-tab}). All the observations were made through a broad BVR filter with a roughly-square passband from 4130~\AA\ to 7385~\AA. We reduced the data using standard IRAF routines and extracted the brightness of V1408 Aql relative to the same two comparison stars used by \cite{bay11}.

We combined the new photometry with the previously published photometry obtained using the same equipment and reduced in a similar manner by \cite{bay11} and \cite{mas12}. Altogether we now have 29 nights of data extending from 2008 to 2013. The six-year light curve is plotted in Figure~\ref{alldata-fig}. On the scale of this figure the individual nights are not resolved, only entire observing runs. The data in the four clumps on the left side of the figure were obtained by \cite{bay11} and \cite{mas12}, while the new data comprise the two clumps on the right side of the figure. The mean brightness of V1408 Aql varies by a factor of two. The night to night variations are of the order of one day, while the rapid variations or flickering are occasionally observed approximately every $ 1.1 \pm 0.2 $ hours. These long term variations might be caused by changes in the accretion rate of the secondary, which in turn affect the emission from the accretion disk. The detached clump of bright points on the right side of Figure \ref{alldata-fig} comes from a single night, 14 July 2012, when V1408 Aql was $\sim50$\% brighter than the immediately following night. Figure \ref{runs-fig} shows just the nine new light curves. The variation in mean brightness from night to night is readily apparent as is the sinusoidal orbital modulation. Other sources similar to V1408 Aql show flares and bursts. For example, the low mass black hole LMXB GRO J0422+32 was found to have episodic gamma-ray and neutrino emission observed in the form of a hard power-law in the X-ray spectra \citep{vie12}.

\begin{figure}
	\centering
	\subfigure[June 06, 2008]
	{
		\includegraphics[scale=0.18]{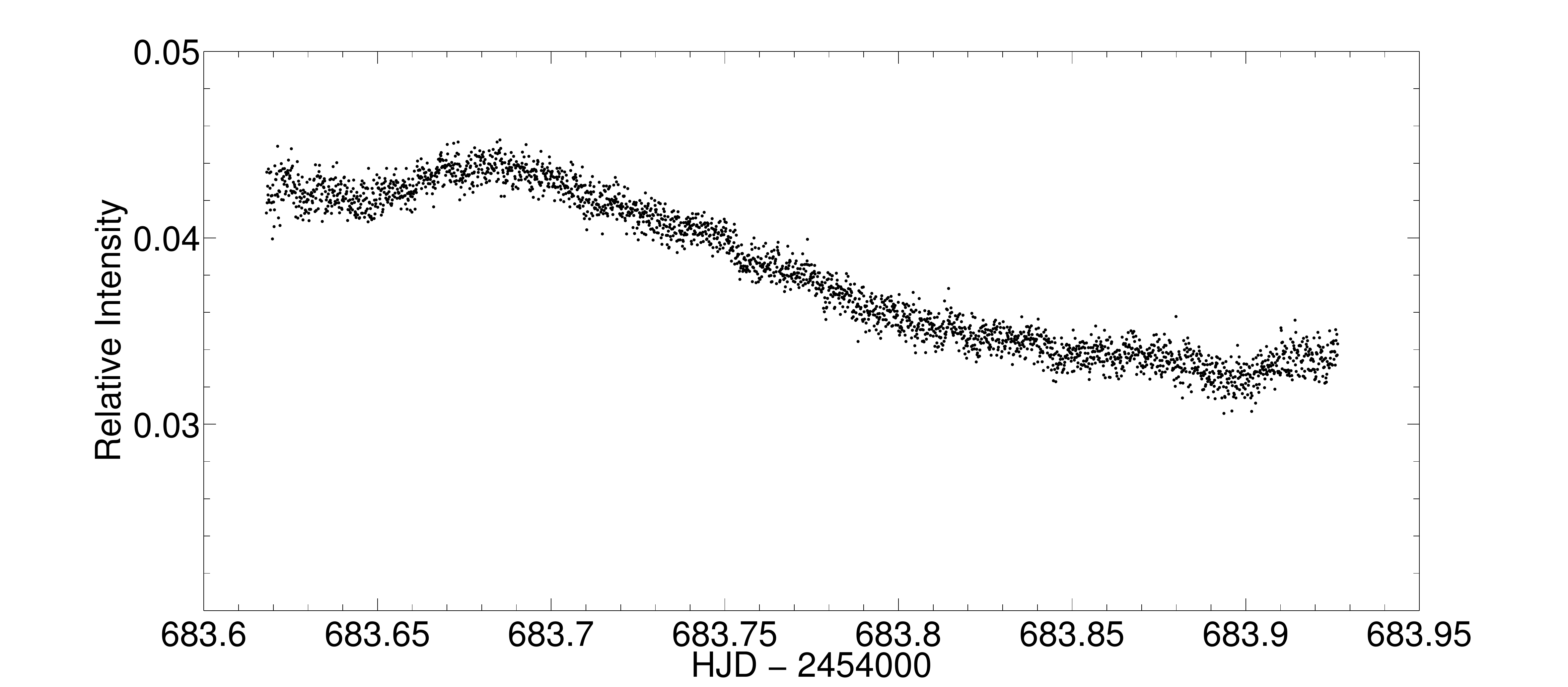}
		\label{ronea}
	}
	\\
	\subfigure[June 25, 2009]
	{
		\includegraphics[scale=0.18]{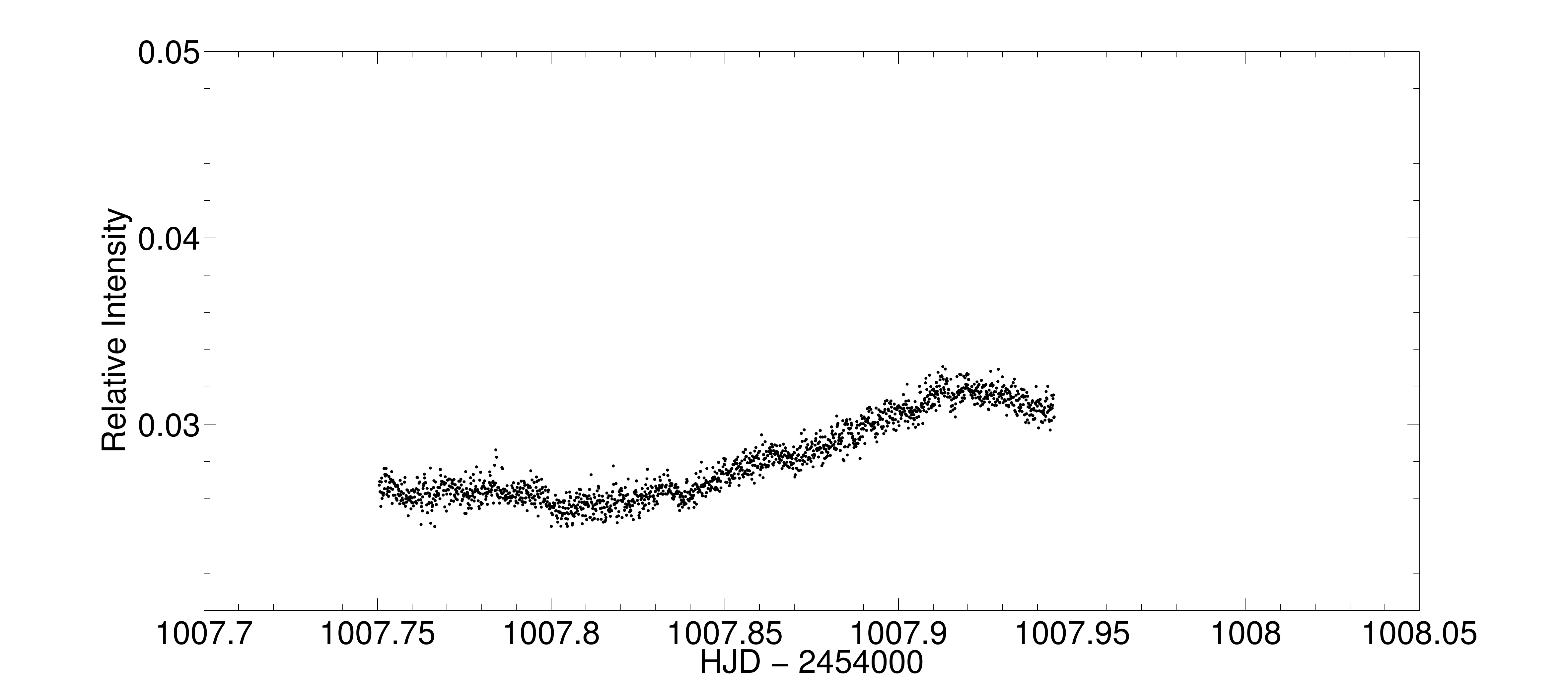}
		\label{roneb}
	}
	\caption{Two light curves of V1408 Aql. 
		On June 06, 2008 (top) V1408 had a high mean brightness and the peak-to-peak
		amplitude of its orbital modulation was 29\%. 
		On June 25, 2009 (bottom) its mean brightness was lower and the amplitude 
		of the modulation was 22\%.}
	\label{runs2-fig}
\end{figure}

\begin{figure}
	\begin{center}
		\includegraphics[angle=0.0,scale=0.20]{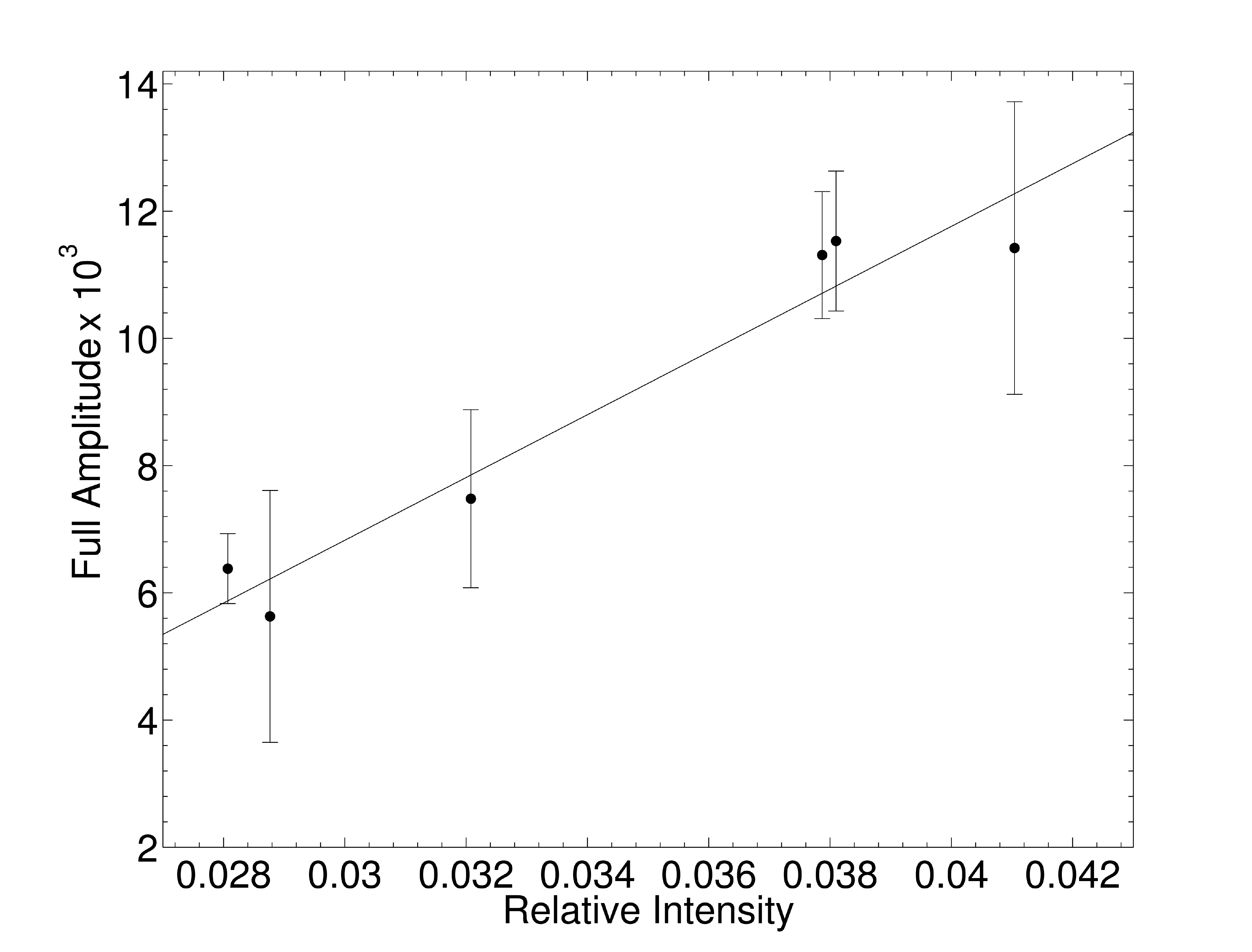}
	\end{center}
	\caption[New 2008 light curve.]{The peak-to-peak amplitude of the orbital modulation
		as a function of mean brightness from the six nights for which we have enough data
		to measure both accurately.
		The amplitude of the modulation increases as the mean brightness increases.}
	\label{amp-fig}
\end{figure}

The amplitude of the sinusoidal modulation is correlated with the mean brightness. Figure~\ref{ronea} shows a light curve from a night when V1408 Aql was unusually bright and Figure~\ref{roneb} shows a light curve when it was unusually faint. The sinusoidal modulation is clearly visible on both nights but its peak-to-peak amplitude was 29\% on the bright night and only 22\% on the faint night. Figure~\ref{amp-fig} shows the amplitude of the sinusoidal modulation plotted against mean brightness from the six nights for which we have enough data to measure both accurately. There is a nearly linear correlation between the two with a slope $dA/dB = 0.49$, where $A$ is the peak-to-peak amplitude of the sinusoidal modulation and $B$ is the mean brightness, both $A$ and $B$ are in units of relative intensity.

\section{The Orbital Ephemeris}
 
\begin{figure}
	\begin{center}
		\includegraphics[angle=0.0,scale=0.17]{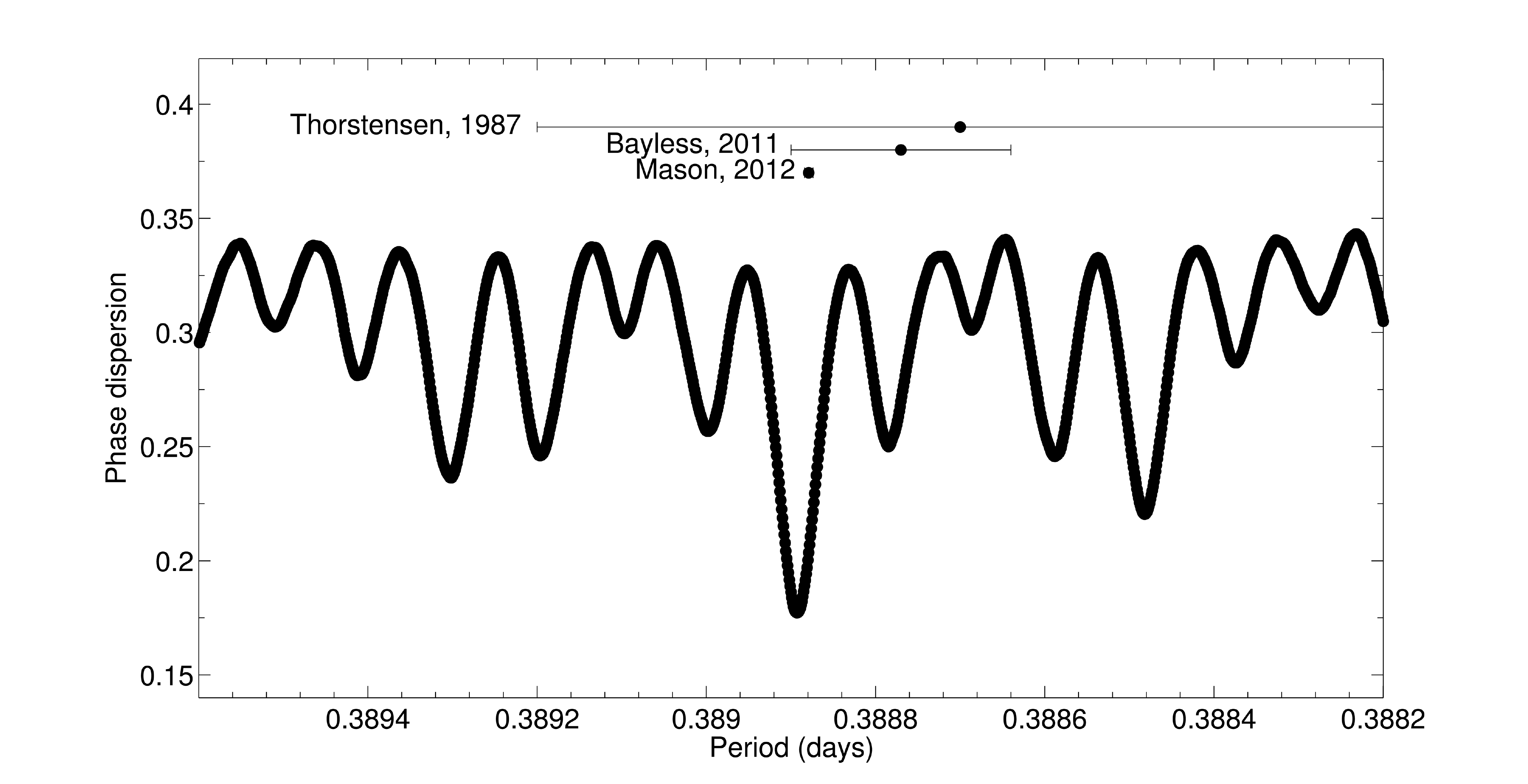}
	\end{center}
	\caption[New 2008 light curve.]{The PDM periodogram for all our photometry of V1408 Aql. 
		The lowest minimum is the best orbital period. 
		The points above the periodogram mark the previously published orbital periods from 
		\citet{tho87}, 
		\citet{mas12}, and \cite{bay11}.}
	\label{PDM-fig}
\end{figure}

The night to night variations of the mean brightness tend to obscure the orbital modulation of the light curve, adding noise to measurements of the orbital period. The long term variations are on the order of days, or over twice as long as the orbital period. To mitigate this problem we multiplied the relative brightness on each night by a normalization factor to scale all the light curves to the same mean brightness, which we chose to be near the minimum mean brightness. There are a few caveats to mention. This works to first order because the long term variations are longer than the orbital period. And even though each light curve does not cover exactly the same phase, the normalization method for scaling should not be a problem. Since the normalization value used was less than 10\% for more than 50\% of the light curves used. We then measured the orbital period using the Phase Dispersion Minimization (PDM) periodogram \citep{ste78}. Figure~\ref{PDM-fig} shows the PDM periodogram along with the previously published periods and their error bars. The minimum of the periodogram is at a period of 0.388893(3) days, which is consistent with, but more accurate, than previously published periods \citep{tho87,bay11,mas12}. The improved orbital ephemeris is:
\begin{equation}
T = {\rm HJD}\, 2454621.829(4) + 0.388893(3) E,
\end{equation}
where $T$ is the time of maximum flux and $E$ is the orbit number. The five-year span of our data is too short for a meaningful constraint on the rate of change of the orbital period. 

\begin{figure}
	\begin{center}
		\includegraphics[angle=0.0,scale=0.068]{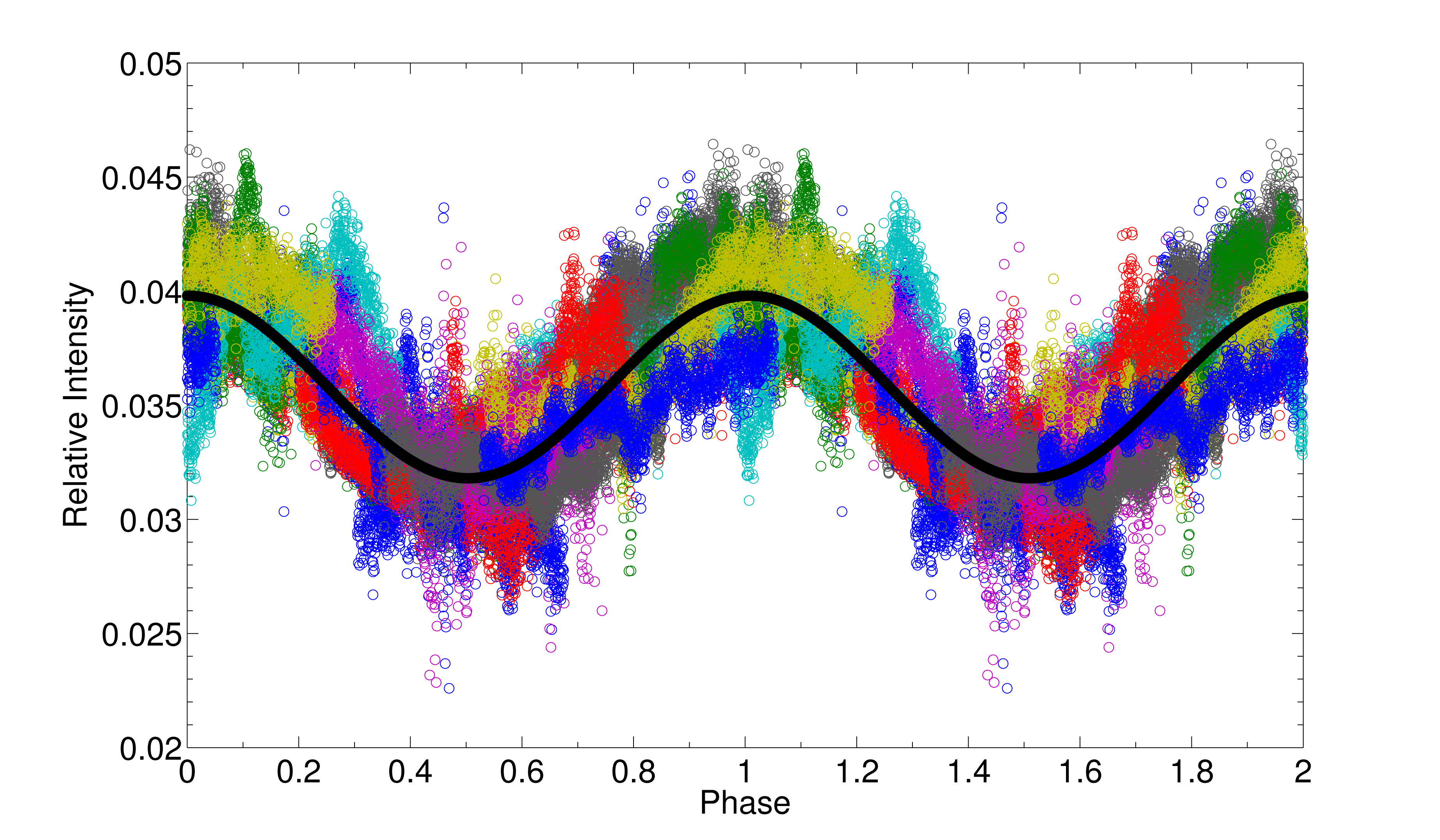}
	\end{center}
	\caption[New 2008 light curve.]{Scaled photometry of V1408~Aql from all 29 nights folded at the orbital period. 
		Each color represents a separate night.
		Phase zero is defined to be the maximum of the mean orbital light curve. 
		The solid black line is the best-fit sine curve, not a model of the modulation.}
	\label{phase3-fig}
\end{figure}

Figure \ref{phase3-fig} shows the scaled data from all 29 nights folded at the orbital period. The sinusoidal orbital modulation is apparent, and so too is a large scatter about the mean orbital variation. The scatter is caused by rapid variations - flickering - in the optical flux, which we presume is caused by rapid variations in the flux from the accretion disk. Figure~8 of \citet{mas12} plots nightly light curves of V1408~Aql at a scale that displays the flickering particularly well. The average orbital light curve is shown in Figure~\ref{phase3-fig}.

\section{Analysis of the Orbital Light Curve}

We analyzed the orbital light curve of V1408 Aql with our \texttt{XRbinary} light curve synthesis program. All the models consisted of a black hole primary star surrounded by an accretion disk, plus a secondary star that fills its Roche lobe and is irradiated by flux from the disk. For our purposes the black hole is just a point source of gravity that emits no flux. The accretion disk is a cylindrically-symmetric, geometrically thin, optically thick, steady-state alpha-model disk. The outer radius of the disk was set to the tidal truncation radius, which we take to be 0.9 times the mean radius of the primary star's Roche lobe \citep{fra02}. We assume that the disk radiates like a black body.

The flux emitted from a surface element of the secondary star is prescribed to be $F_{emit} \ = \ F_0 + \alpha F_{irr}$, where $F_0$ is the flux the secondary star would emit in the absence of irradiation, $F_{irr}$ is the irradiating flux from the accretion disk, and $\alpha$, which we call the ``albedo,'' is the fraction of $F_{irr}$ that is re-radiated instead of absorbed into the structure of the secondary star. The intrinsic flux is calculated from the gravity darkening law $F_0 \propto |g|^{4\beta}$, where the local gravity is determined from the Roche geometry, and $\beta$ is the temperature-dependent gravity-darkening coefficient, which we take from Claret (2000). We assume the emitted flux is fully thermalized, so the local effective temperature is given by $\sigma T_{eff}^4 \ = \ F_{emit}$. Armed with a local effective gravity and local effective temperature, we adopt Kuruz solar-composition spectra for the local emitted flux if $T_{eff} \leq 8000$ K and blackbody spectra otherwise. The model does not include spots or other features on the surface of the secondary star.

The free parameters of this model are
\vspace{-0.75\baselineskip}
\begin{itemize}
   \item  the masses of the primary and secondary stars.
   \vspace{-0.5\baselineskip}
   \item the orbital inclination.
      \vspace{-0.5\baselineskip}
   \item the inner radius and total luminosity of the accretion disk.
      \vspace{-0.5\baselineskip}
   \item the intrinsic effective temperature of the (unirradiated) secondary star.
      \vspace{-0.5\baselineskip}
   \item  the albedo of the secondary star.
\end{itemize}
   \vspace{-0.75\baselineskip}
There are also some additional parameters such as the time of phase zero and the resolution of the time steps that have no effect on the structure of the system.

\begin{figure}
	\begin{center}
		\includegraphics[angle=0.0,scale=0.3]{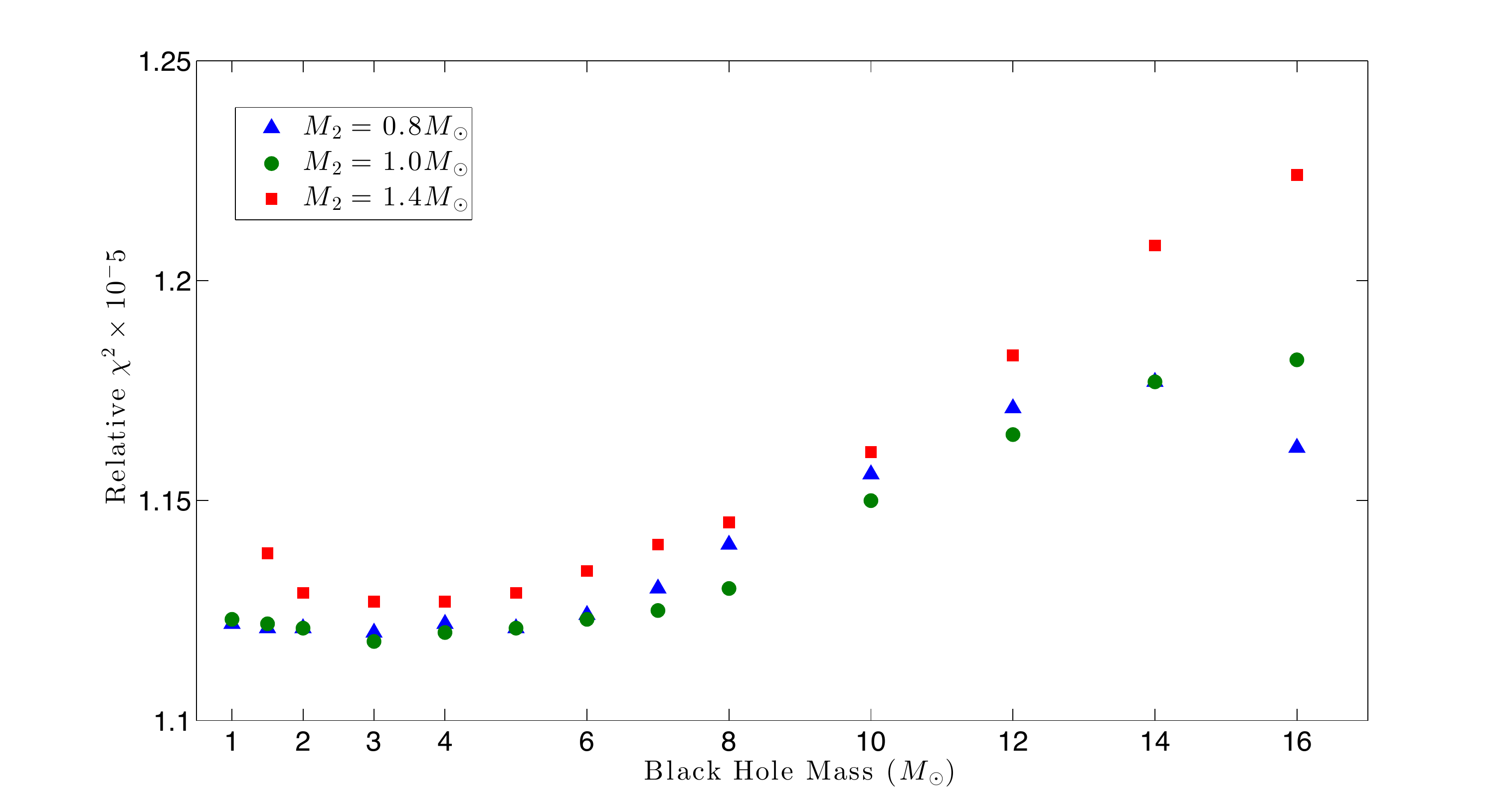}
	\end{center}
	\caption[Mass Vs Chi^2]{Plot of the relative $\chi^2$ for the best-fitting synthetic light
		curves as a function of $M_1$.
		The three sets of points correspond to the three masses of $M_2$. We see that the best models are for a primary of $3M_{\odot}$ for any secondary mass.}
	\label{mass}
\end{figure}

Perhaps surprisingly, our results do not depend strongly on the albedo of the secondary star. Changes in the value of the albedo are absorbed almost entirely into offsetting changes in the luminosity of the accretion disk. We have, therefore, fixed the albedo at 0.5. To estimate the inner radius of the disk we assumed the black hole has a spin of $a^* = 0.9$ and used the spin-ISCO relation in Figure 2 of \citep{mc11}. In fact, though, our results depend only weakly on the inner radius of the disk. The inner radius does affect the fraction of disk flux emitted at optical wavelengths, but even here it is the outer disk radius that dominates the amount of optical flux. A numerical test showed that increasing the inner disk radius by a factor of three increased the preferred orbital inclination by only two degrees. The intrinsic temperature of the secondary star is nearly irrelevant since the optical flux is dominated by the accretion disk and the secondary's heated face. Despite the large number of parameters needed to model the light curve, in the end only four parameters really matter: the mass of the primary star, the mass of the secondary star, the orbital inclination, and the disk luminosity.

The light curve synthesis code assumes that the heating of the secondary star is due to incident flux on it's surface and that this flux is fully thermalized in the atmosphere of the secondary star. Since the incident flux is fully thermalized, the details of the actual spectrum become irrelevant in favor of the total amount of incident energy from the accretion disk. Therefore, the total flux depends on the total luminosity of the disk, not its spectral energy distribution. The disk luminosity is set via a separate input parameter in the code.

The code assumes that the the spectrum emitted by an area element on the surface of the secondary is the same as the spectrum that would be emitted by an isolated star with the same effective temperature. The code does not account for the fact that the regions of the secondary's atmosphere which are heated by radiation can have a temperature inversion, which in turn can alter the absorption-line spectrum and produce chromospheric emission lines \citep{bar04, waw09}. But since we are doing this analysis with optical photometry instead of spectroscopy, the form in which the code handles the incident radiation is not expected be the limiting factor in our results.

\begin{figure}
	\centering
	\subfigure[The best-fitting synthetic light curve for an inclination of $75\,^{\circ}$. 
	The dip near phase zero is caused by an eclipse of the secondary star by the accretion disk.]
	{
		\includegraphics[scale=0.37]{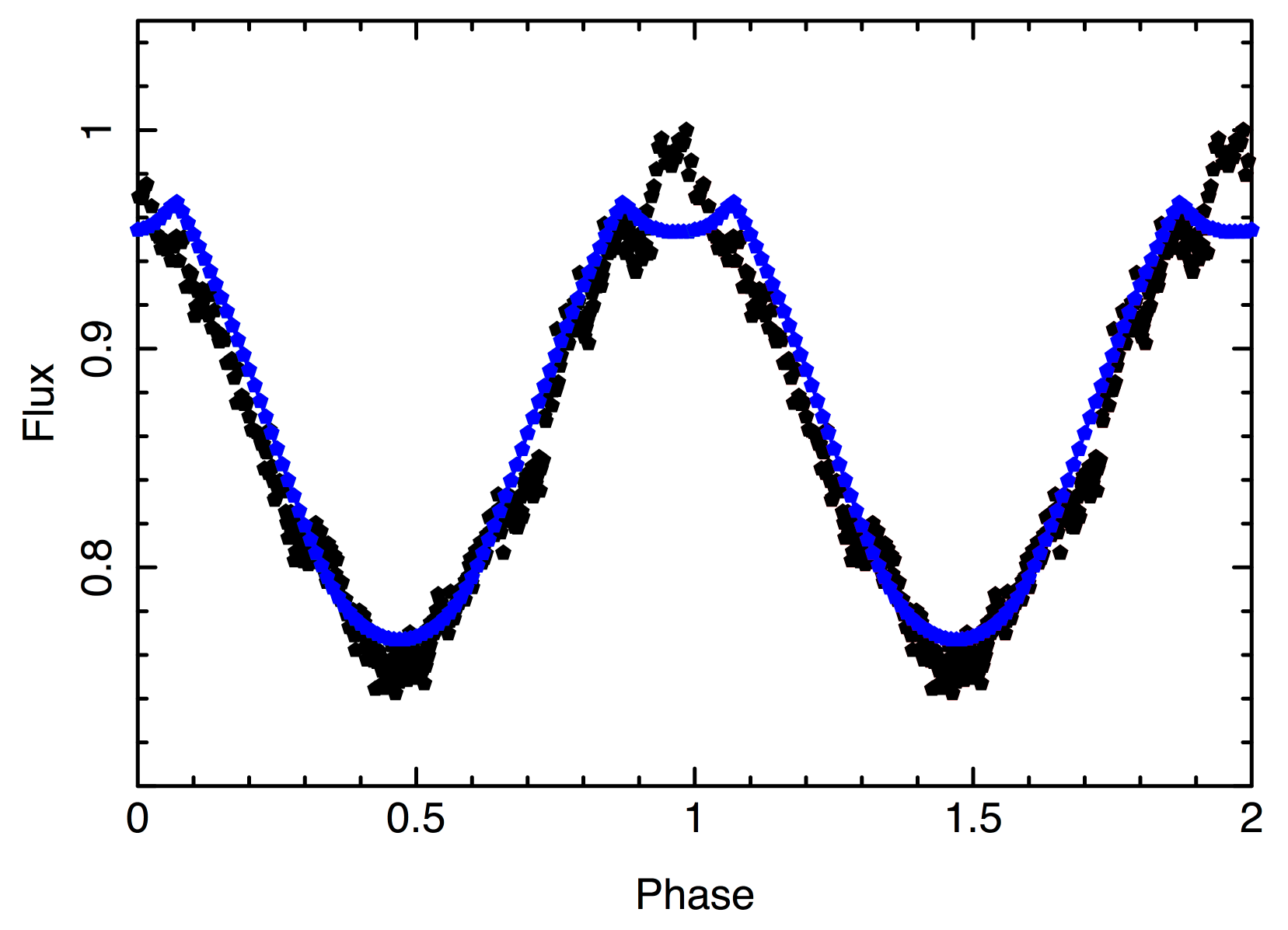}
		\label{bad}
	}
	\\
	\subfigure[The very best fitting synthetic light curve. 
	The parameters of the model are $M_1 = 3.0 M_\odot$, $M_2 = 1.0M_\odot$, and $i = 12.75^{\circ}$]
	{
		\includegraphics[scale=0.37]{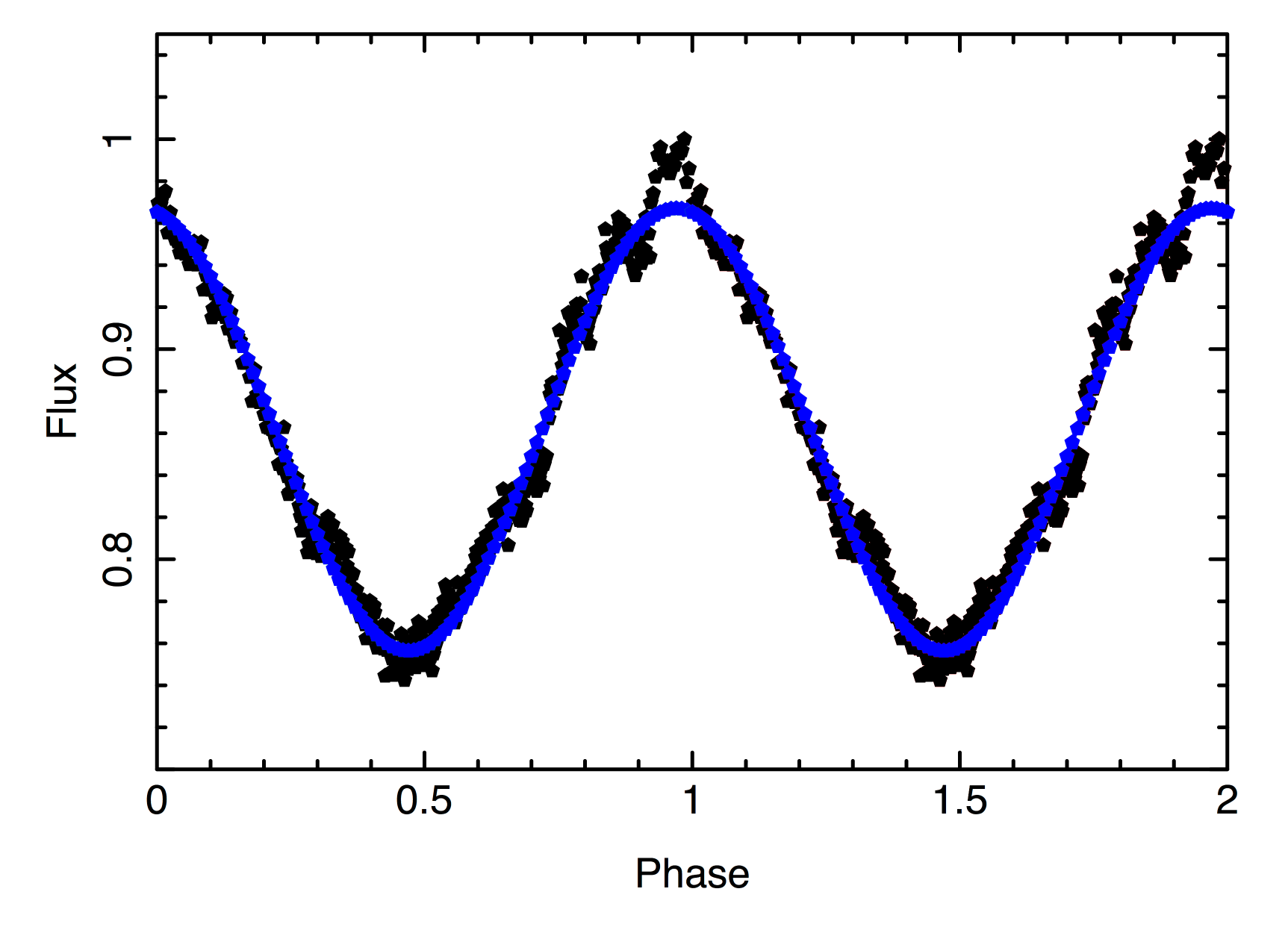}
		\label{good}
	}
	\caption{Two best-fit synthetic light curves overplotted on the mean orbital light curve of V1408 Aql. There is a difference between the average light curve and the model fits at phase = 1.0. This is likely due to insufficient coverage at this phase. However, additional photometry is needed to confirm this result.}
	\label{fits-fig}
\end{figure}

We can eliminate inclinations greater than $65^\circ$. At inclinations greater than $i \approx 65^\circ$ the accretion disk eclipses the secondary star. This is the case for all of the secondary masses modeled ($M_2 = 0.8 - 1.4\, M_\odot$ ). Figure~\ref{bad} shows an example of the distinctive eclipse profile produced when the accretion disk passes in front of the secondary star. No eclipse of this kind has been observed.\footnote{If the disk eclipses the secondary star, the secondary star also eclipses the disk 1/2 orbit later, but a shallow disk eclipse can be difficult to discern because it is superimposed on the minimum of the sinusoidal orbital variation.} To avoid an eclipse at $75^\circ$ either the outer radius of the accretion disk would have to be unrealistically small, about 50\% the radius of the black hole's Roche Lobe, or the mass ratio would have to be so low that the fits to the orbital light curve become unacceptably poor. 

\begin{figure}
	\begin{center}
		\includegraphics[angle=0.0,scale=0.066]{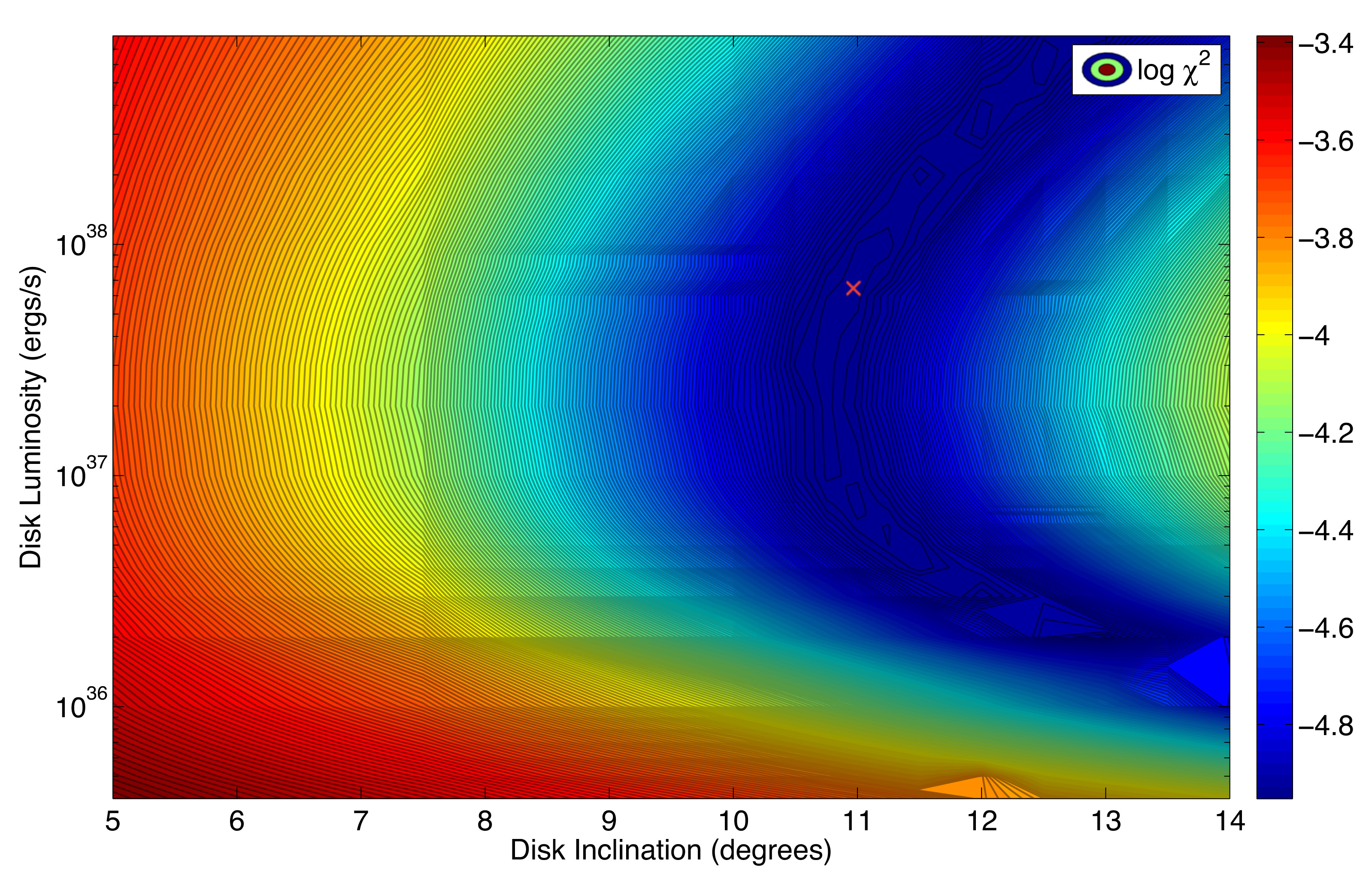}
	\end{center}
	\caption[Fixed inclination models]{The contour plot represents various models with different inclinations and disk luminosities for the case of a $2 M_\odot$ primary and a $0.8M_\odot$ secondary. The red regions represent the worst fits and the blue regions represent the lowest $\chi^2$ and therefore the best fits to the optical light curve. The red cross marks the spot of the best fit with an inclination of $i  = 11\,^{\circ}$ and a luminosity of L = 7.50\e{37} ergs/s.}
	\label{contour}
\end{figure}

Because so few parameters dominate the model, we were able to calculate a grid of models that covered the likely parameter space. The grid points in the primary mass were set at intervals 1 or 2 solar masses between $2\, M_\odot$ to $16\, M_\odot$ (see the first column of Tables 2, 3, and 4).  The upper limit to the mass of a main-sequence secondary star that just fills the Roche lobe is $\sim 1.4\, M_\odot$ \citep{pat05}. We therefore chose grid points in secondary mass at $1.4\, M_\odot$, $1.0\, M_\odot$, and $0.8\, M_\odot$, the latter two corresponding to evolved secondaries. The grid points in inclination were typically at half degree intervals and in disk luminosity at intervals of roughly 10\% of the luminosity. For the models with a secondary star mass of $1.4\ M_\odot$ we adopted a temperature of 6650~K, which is appropriate for a main-sequence star of that mass. For the models with secondary masses of $0.8\, M_\odot$ and $1.0\, M_\odot$ we adopted temperatures of 4700~K and 4900~K respectively, which are more appropriate for an  evolved secondary. Again, though, the intrinsic temperature of the secondary has little effect on the synthetic light curve. The output at each grid point is a synthetic light curve and a value of relative $\chi^2$ for the fit of the synthetic light curve to the observed data. We report a relative $\chi^2$ because the noise in the light curves is dominated by flickering noise, which is both highly correlated and variable in time, making measurement of the absolute $\chi^2$ intractable. The relative values of $\chi^2$ are, though, adequate for determining the best-fit parameters and their standard deviations. Plots like the one shown in Figure \ref{contour} were used to pinpoint the inclination and disk luminosity that yield the smallest relative $\chi^2$ for each pair of masses. In that Figure, the red zones have the highest values of $\chi^2$, while the blue regions have the lowest values and correspond to better fits to the data. Once the masses have been chosen, the inclination is highly constrained, typically to within $0.5^\circ$, the disk luminosity less so, but still typically to within 30\%.

The results are given in Tables~\ref{table:eight}, \ref{table:one}, and \ref{table:four}, which list the relative $\chi^2$ for the the best-fit values of the inclination and disk luminosity for each pair of primary and secondary masses. Figure~\ref{mass} intercompares the models by plotting the best-fit values of $\chi^2$ against the mass of the primary star. The lowest values of $\chi^2$ occur for $M_2 = 0.8$ and $1.0\, M_\odot$, and for $M_1$ between 2 and $5\, M_\odot$. The 90\% upper bound on the mass of the black hole is $6.2\, M_\odot$. We are unable to place a useful lower limit on the mass of the black hole, but masses in the range of the known neutron star masses are not excluded. The very best fit occurs for $M_1 = 3.0\, M_\odot$, $M_2 = 1.0\, M_\odot$, and an orbital inclination $i = 12.75^{\circ}$. The preferred mass of the secondary in our models is perfectly consistent with the results found by \cite{hak14}. Our models show a 90\% upper bound of $6.2\, M_\odot$, and a 90\% lower bound that is outside the reasonable mass for a black hole primary. The $1 \sigma$ deviation from the best fit of $M_1 = 3.0\, M_\odot$ is of $\pm 2.5\, M_\odot$. Even though we can not constrain the lower mass based solely on our modeling, we adopt a lower bound of $M_1 = 2.0\, M_\odot$ from the assumption that the primary is a black hole and not a neutron star. Black holes are expected to form only when the mass of the compact object is too high to become a neutron star. Some ways in which a black hole can be created include: the collapse of the core of a star during a core-collapse supernova, a merger of two neutron stars, and the collapse of a neutron star that is accreting mass. In all these scenarios the black hole will not form, unless the mass is greater than the highest possible mass for a neutron star. Currently, the observed upper mass limit of neutron stars is $2.0\, M_\odot$ \citep{dem10}. Figure \ref{good} shows the synthetic light curve for this model overplotted on the mean optical light curve. The correlated errors in the light curves and imperfect model physics broaden the range of permitted black hole masses beyond the formal 90\% upper limit, possibly even to much higher masses. The best-fit masses are, however, unchanged by these considerations.

\begin{table*}[ht] 
	\caption{Models with a Secondary Star Mass of $0.8M_\odot$} 
	\centering 
	\begin{center}
		\begin{tabular}{l*{8}{c}r}
			\hline \hline
			
			$M_1$    &     $a$      & $R_{Disk}$ &    $i$     &              $L_{D}$              & $L_{D}/L_{Edd}$ & Relative $\chi^2$  \\
			($M_\odot$) &    (AU)     &         (a)       &  (deg)   & ($10^{38}$ erg s$^{-1}$) &                             &        ($\e{5}$)         \\
			\hline
			2         &   0.0145   &      0.414     &   11.0    &                  0.75                &           0.30          &        1.121        \\
			3         &   0.0160   &      0.446     &   13.5    &                  0.72                &           0.19          &        1.120        \\
			4         &   0.0173   &      0.469     &   16.0    &                  0.71                &           0.14          &        1.122        \\
			5         &   0.0184   &      0.486     &   18.5    &                  0.78                &           0.12          &        1.121        \\
			6         &   0.0194   &      0.499     &   21.0    &                  0.70                &           0.09          &        1.124        \\
			7         &   0.0204   &      0.511     &   23.5    &                  0.72                &           0.08          &        1.130        \\
			8         &   0.0212   &      0.520     &   26.0    &                  0.75                &           0.07          &        1.140        \\
			10         &   0.0227   &      0.536     &   31.0    &                  0.90                &           0.07          &        1.156        \\
			12         &   0.0240   &      0.548     &   36.0    &                  1.30                &           0.09          &        1.171        \\
			14         &   0.0252   &      0.558     &   41.0    &                  1.60                &           0.09          &        1.177        \\
			16         &   0.0263   &      0.567     &   46.0    &                  2.90                &           0.14          &        1.162        \\
			\hline
		\end{tabular}
	\end{center}
	\raggedright
	
	Note: $a$ is the separation of the stars, $R_{Disk}$ is the outer radius of the disk in units of $a$,
	$L_{D}$ is the luminosity of the disk, and $L_{Edd}$ is the Eddington luminosity.
	\label{table:eight} 
\end{table*} 

\begin{table*}[ht] 
	\caption{Models with a Secondary Star Mass of $1.0M_\odot$} 
	\centering 
	\begin{center}
		\begin{tabular}{l*{8}{c}r}
			\hline \hline
			
			$M_1$    &     $a$      & $R_{Disk}$ &    $i$     &              $L_{D}$              & $L_{D}/L_{Edd}$ & Relative $\chi^2$  \\
			($M_\odot$) &    (AU)     &         (a)       &  (deg)   & ($10^{38}$ erg s$^{-1}$) &                             &        ($\e{5}$)         \\
			\hline
			2         &   0.0148   &      0.396     &   10.50  &                  1.4                  &           0.56          &        1.121        \\
			3         &   0.0163   &      0.429     &   12.75  &                  1.3                  &           0.34          &        1.118        \\
			4         &   0.0176   &      0.451     &   15.00  &                  1.4                  &           0.28          &        1.120        \\
			5         &   0.0187   &      0.469     &   17.25  &                  1.4                  &           0.22          &        1.121        \\  
			6         &   0.0196   &      0.482     &   19.50  &                  1.6                  &           0.21          &        1.123        \\
			7         &   0.0205   &      0.494     &   21.75  &                  1.6                  &           0.18          &        1.125        \\
			8         &   0.0214   &      0.504     &   24.00  &                  1.6                  &           0.16          &        1.130        \\
			10         &   0.0228   &      0.520     &   28.50  &                  1.9                  &           0.15          &        1.150        \\
			12         &   0.0241   &      0.533     &   32.50  &                  1.7                  &           0.11          &        1.165        \\
			14         &   0.0253   &      0.544     &   36.50  &                  1.7                  &           0.10          &        1.177        \\
			16         &   0.0264   &      0.552     &   41.00  &                  2.8                  &           0.14          &        1.182        \\
			
			\hline
		\end{tabular}
	\end{center}
	\raggedright

	See note to Table~\ref{table:eight}.
	\label{table:one} 
\end{table*} 

\begin{table*}[ht] 
	\caption{Models with a Secondary Star Mass of $1.4M_\odot$} 
	\centering 
	\begin{center}
		\begin{tabular}{l*{8}{c}r}
			\hline \hline
			
			$M_1$    &     $a$      & $R_{Disk}$ &    $i$     &              $L_{D}$              & $L_{D}/L_{Edd}$ & Relative $\chi^2$  \\
			($M_\odot$) &    (AU)     &         (a)       &  (deg)   & ($10^{38}$ erg s$^{-1}$) &                             &        ($\e{5}$)         \\
			\hline
			2         &   0.0154   &      0.369     &   10.0   &                   1.2                  &           0.48          &        1.129        \\
			3         &   0.0168   &      0.401     &   12.0   &                   1.7                  &           0.45          &        1.127        \\
			4         &   0.0180   &      0.425     &   14.0   &                   1.9                  &           0.38          &        1.127        \\
			5         &   0.0191   &      0.442     &   16.0   &                   2.3                  &           0.37          &        1.129        \\
			6         &   0.0200   &      0.457     &   18.0   &                   2.8                  &           0.37          &        1.134        \\
			7         &   0.0209   &      0.469     &   20.0   &                   3.0                  &           0.34          &        1.140        \\
			8         &   0.0217   &      0.479     &   22.0   &                   3.4                  &           0.34          &        1.145        \\
			10         &   0.0231   &      0.496     &   26.0   &                   4.2                  &           0.33          &        1.161        \\
			12         &   0.0244   &      0.509     &   30.0   &                   5.0                  &           0.33          &        1.183        \\
			14         &   0.0256   &      0.520     &   34.0   &                   6.3                  &           0.36          &        1.208        \\
			16         &   0.0266   &      0.530     &   38.0   &                   8.9                  &           0.44          &        1.224        \\
			\hline
		\end{tabular}
	\end{center}
	\raggedright

	See note to Table~\ref{table:eight}.
	\label{table:four} 
\end{table*}

\section{Discussion}

The fits to the orbital light curve of V1408~Aql favor low orbital inclinations, much lower than the inclinations usually adopted in analyses of its X-ray spectral energy distribution. According to \cite{mai14}, for example, fits to the X-ray spectrum of 4U~1957+115 require an orbital inclination near $75^\circ$ to obtain the canonical X-ray spectral hardening factor $h_d = T_{color}/T_{eff} = 1.7$ \citep{dav05,dav06} (this quantity is often called the color correction factor and denoted by $f_{c}$). Lower inclinations yield values for $h_d$ that are greater than 1.7. To rigidly limit the color correction factor to values near 1.7 is, however, unwarranted. Authors such as \cite{now08} preferred high inclinations near $75^\circ$, which in combination with a large distance, low black hole mass and high accretion rate reconcile the high temperature and low normalization obtained from their models. For some of their models they fixed the inclination to $75^\circ$ in order to be consistent with the interpretation of optical observations of \cite{hak99}. There is ample theoretical and observational evidence that color correction factors can be much larger than 1.7, Cygnus X-1 shows one of the highest color correction factors of up to $f_c \sim 5 $ \citep{rey13}. Color correction factors are different for various X-ray sources, and beyond that, an individual source can have different color correction factors \citep{mer00,dun11,sal13}.

While it is not our intent to re-analyze the X-ray spectral energy distribution, we note that color correction factors larger than 1.7 do permit reasonable system  parameters for low orbital inclinations. Although the typical value for the spectral hardening factor is $f_c = 1.7 $, \cite{mai14} provided a value of $f_c = 2.0 - 2.2$ for their preferred model. \citet{now12} fit the spectral energy distribution of of 4U~1957+115 with several sets of models, one of which was the {\tt eqpair} model with color correction factors raging from $f_c = 1.7 - 3.3$. While the fits of the {\tt eqpair} model yielded acceptable values for $\chi^2$, they required a low value for the normalization factor, $N_{eqp} = 1.926\e{-4}$. \citet{now12} interpreted the low normalization factor as evidence for a small inner disk radius and, consequently, evidence for a rapidly spinning black hole. While the fits all assumed an orbital inclination of $75^\circ$, the parameters of the fits are highly degenerate. \citet{now12} give a scaling relation for other system parameters:
\begin{equation}
   N_{eqp} = \left(  \dfrac{M_1}{1\, M_\odot}     \right)^2  
             \left(  \dfrac{D}{1\, \textrm{kpc}} \right)^{-2}
              f_{c}^{-4} \cos i,
\end{equation}
where $D$ is the distance and $f_c$ is the color correction factor. 
With $M_1 = 3\, M_\odot$, $i = 12.75^\circ$, and $N_{eqp} = 1.926\e{-4}$, this becomes
\begin{equation}
   D f_c^2 = 213.5.
\end{equation}
If one insists on $f_c = 1.7$, the distance would be $D = 74\ \textrm{kpc}$, placing V1408~Aql uncomfortably far outside the Galaxy. If, instead, we set $f_c = 2.3$, the distance becomes 40~kpc; and if $f_c = 3.3$, it drops to 20~kpc, either of which would place V1408~Aql in the halo of the Galaxy.

This range of distances agrees with the results of \citet{rus10}. If V1408~Aql contains a black hole, it would need to be at a distance between 22 and 40~kpc to lie in the same region of the $(L_X,L_{OPT})$ diagram as other black hole X-ray binaries in their soft states, with the larger distance somewhat preferred (see Figure~8 in \citet{rus10}). At the larger distance the X-ray luminosity at 2-10~keV would be $\sim\! 10^{38}\ \textrm{erg s}^{-1}$. This agrees with the large, distance-independent disk luminosity that is required to heat the secondary star in our models of the orbital light curve.

Finally, we have found a correlation between the mean brightness of V1408~Aql and the amplitude of the sinusoidal orbital modulation:  The amplitude increases as the mean brightness increases (see Figure~\ref{amp-fig}). \citet{rus10} found a correlation between the mean optical and mean X-ray fluxes from V1408~Aql:  The mean X-ray flux increases as the mean optical flux increases. Taken together, these two correlations imply that the amplitude of the sinusoidal orbital modulation increases as the X-ray flux increases. This is the correlation expected from a model in which the orbital modulation is caused by the heated face of the secondary star. The X-ray flux is a good proxy for the total disk luminosity. As the X-ray flux increases, the disk becomes more luminous and heats the face of the secondary star towards the disk to a higher temperature, increasing the amplitude of the orbital variation. 

\section{Summary}

We have presented new optical high-speed photometry of V1408~Aql from nine nights in 2012 July and August. The optical light curve continues to display a nearly-sinusoidal orbital modulation along with night-to-night variations of the mean brightness. We combined the new photometry with our previously-published photometry to derive a more accurate orbital period and mean orbital light curve, and to better define the night-to-night variations. We find that the amplitude of the orbital modulation is strongly correlated with the mean brightness, $dA/dB = 0.49$, where $A$ is the peak-to-peak amplitude of the sinusoidal modulation and $B$ is the mean brightness, both $A$ and $B$ are in units of relative intensity. The relative amplitude of the orbital modulation rises from 23\% when V1408~Aql is at the minimum of the observed range of its  brightness to 29\% at the maximum of the range. We attribute the changes in mean brightness to changes in the luminosity of the accretion disk around the black hole.

After scaling all the nightly light curves to the same mean brightness, we derived a more accurate orbital period,  0.388893(3) days, and mean orbital light curve, shown on Figure~\ref{good}. The mean orbital light curve is consistent with a model in which the orbital modulation is caused entirely by the changing aspect of the heated face of the secondary star as it revolves around the black hole. Fits of synthetic orbital light curves based on this model to the observed light curve favor low orbital inclinations and low black hole masses, the best fit occuring for $M_1 = 3.0\, M_\odot$, $M_2 = 1.0\, M_\odot$, and $i=12.75^{\circ}$. The upper bound to the mass of the black hole is $6.2\, M_\odot$ with a 90\% probability, although uncertainties in the data and the models allow higher masses, possibly much higher masses. Orbital inclinations higher than about $65^\circ$ are strongly disfavored by the lack of eclipses.

The low orbital inclinations we have found are compatible with previous analyses of the X-ray spectral distribution of V1408~Aql if the color correction factor is somewhat larger that the value typically adopted for the analyses. If the distance to V1408~Aql is 40~kpc, the color correction factor must be increased to 2.3; and if the distance is 20~kpc, the color correction factor is 3.3.

In conclusion, the compact star in V1408~Aql a viable candidate for a black hole whose mass lies within the gap in the distribution of compact star masses.

\acknowledgements
We thank Tom Maccarone and Michael Nowak for helpful discussions regarding the nature of the compact object and Amanda Bayless for providing some of the data used for this project. We also thank an anonymous referee for providing helpful suggestions towards the improvement of this paper. This research is supported by NSF Grant No.\ 0958783 and by a MARC Scholarship to the University of Texas at El Paso.

\clearpage


\clearpage

\begin{thebibliography}

\bibitem[Antoniadis et al.(2013)]{ant13} Antoniadis, J. et al. 
\ 2013, Science, 340, 448

\bibitem[Barman et al.(2004)]{bar04} Barman, T.~S., 
Hauschildt, P.~H., \& Allard, F.\ 2004, \apj, 614, 338 

\bibitem[Bayless et al.(2011)]{bay11} Bayless, A.~J., 
Robinson, E.~L., Mason, P.~A., \& Robertson, P.\ 2011, \apj, 730, 43 

\bibitem[Belczynski et al.(2012)]{bel12}
Belczynski, K., Wiktorowicz, G., Fryer, C.\ L., \& Kalogera, V.\ 
2012, \apj, 757, 91

\bibitem[Davis et al.(2005)]{dav05}
Davis, S.\ W., Blaes, O.\ M., Hubeny, I., \& Turner, N.\ J.\ 
2005, \apj, 621, 372

\bibitem[Davis et al.(2006)]{dav06}
Davis, S.\ W., Done, C., \&, Blaes, O.\ M.\
2006, \apj, 647, 525

\bibitem[Demorest et al.(2010)]{dem10} Demorest, P.~B., Pennucci, T.,
Ransom, S.~M., Roberts, M. S. E., \& Hessels, J.~W.,~T.
\ 2010, Nature, 467, 1081

\bibitem[Dunn et al.(2011)]{dun11} 
Dunn, R.\ J.\ H., Fender, R.\ P., K\"ording, E. G., Belloni, T., \& 
Merloni, A.\  
2011, \mnras, 411, 337 

\bibitem[Farr et al.(2011)]{far11} Farr, W.~M., 
Sravan, N, Cantrell, A..~M., et al.\ 2011, \apj, 741, 103 

\bibitem[Filippenko et al.(1999)]{fil99} Filippenko, A.~V., 
Leonard, D.~C., Matheson, T., et al.\ 1999, \pasp, 111, 969 

\bibitem[Frank, King, \& Raine(2002)]{fra02}
Frank, J., King, A., \& Raine, D.\ 2002,
``Accretion Power in Astrophysics'' 
(Cambridge: Cambridge Univ. Press)

\bibitem[Fryer et al.(2012)]{fry12}
Fryer, C.\ L., Belczynski, K., Wiktorowicz, G., Dominik, M.,
Kalogera, V., \& Holz, D.\ E.\ 
2012, \apj, 749, 91

\bibitem[Fryer(1999)]{fry99} Fryer, C.~L.\ 1999, \apj, 522, 
413 

\bibitem[Gelino \& Harrison(2003)]{gel03} 
Gelino, D.\ M., \& Harrison, T.\ E.\ 2003, \apj, 599, 1254 

\bibitem[Giacconi et al.(1974)]{gia74}
Giacconi, R., Murray, S., Gursky, H., et al.\ 
1974, \apjs, 27, 37 
 
\bibitem[Hakala et al.(1999)]{hak99} Hakala, P.~J., Muhli, 
P., \& Dubus, G.\ 1999, \mnras, 306, 701 

\bibitem[Hakala et al.(2014)]{hak14} Hakala, P., Muhli, P., 
\& Charles, P.\ 2014, \mnras, 444, 3802 

\bibitem[Kreidberg(2012)]{kre12}
Kreidberg, L., 
Bailyn, C.~D., Farr, W.~M., \& Kalogera, V.\ 2012, \apj, 757, 36 

\bibitem[Latimer(2014)]{lat14} Latimer, J.~M.
\ 2014, Gen. Rel. and Grav., 46, 1713

\bibitem[Maitra et al.(2013)]{mai14} Maitra, D., Miller, 
J.~M., Reynolds, M.~T., Reis, R., \& Nowak, M.\ 2014, \apj, 794, 85

\bibitem[Margon et al.(1978)]{mar78} Margon, B., Thorstensen, 
J.~R., \& Bowyer, S.\ 1978, \apj, 221, 907 

\bibitem[Mason et al.(2012)]{mas12} Mason, P.~A., Robinson, 
E.~L., Bayless, A.~J., \& Hakala, P.~J.\ 2012, \aj, 144, 108 

\bibitem[McClintock et al.(2011)]{mc11} McClintock, J.~E., 
Narayan, R., Davis, S.~W., et al.\ 2011, Classical and Quantum Gravity, 28, 
114009 

\bibitem[Merloni et al.(2000)]{mer00}
Merloni, A., Fabian, A.\ C., \& Ross, R.\ R.
2000, \mnras, 313, 193

\bibitem[Miller et al.(2011)]{mil11} Miller, J.~M., Miller, 
M.~C., \& Reynolds, C.~S.\ 2011, \apjl, 731, L5 

\bibitem[Nather \& Mukadam(2004)]{nat04}
Nather, R.~E., \& Mukadam, A.~S.\ 2004, \apj, 605, 846 

\bibitem[Nowak et al.(2008)]{now08} Nowak, M.~A., Juett, A., 
Homan, J., et al.\ 2008, \apj, 689, 1199 

\bibitem[Nowak et al.(1999)]{now99} 
Nowak, M.~A., \& Wilms, J.\ 1999, \apj, 522, 476 

\bibitem[Nowak et al.(2012)]{now12} Nowak, M.~A., Wilms, J., 
Pottschmidt, K., et al.\ 2012, American Institute of Physics Conference 
Series, 1427, 48 

\bibitem[{\"O}zel et al.(2010)]{oze10} {\"O}zel, F., Psaltis, 
D., Narayan, R., \& McClintock, J. E. \ 2010, \apj, 725, 1918

\bibitem[Patterson et al.(2005)]{pat05}
Patterson, J.\ et al.\ 2005. \pasp, 117, 1204

\bibitem[Reynolds et al.(2013)]{rey13}
Reynolds, M.\ T., \& Miller, J.\ M.
\ 2013, \apj, 769, 16 

\bibitem[Russell et al.(2010)]{rus10} Russell, D.~M., Lewis, 
F., Roche, P., et al.\ 2010, \mnras, 402, 2671 

\bibitem[Russell et al.(2011)]{rus11} Russell, D.~M., 
Miller-Jones, J.~C.~A., Maccarone, T.~J., et al.\ 2011, \apjl, 739, L19 

\bibitem[Salvesen et al.(2013)]{sal13}
Salvesen, G., Miller, J.\ M., Reis, R.\ C., \& Begelman, M.\ C.\ 
2013, \mnras, 431, 3510 

\bibitem[Singh et al.(1994)]{sin94} Singh, K.~P., Apparao, 
K.~M.~V., \& Kraft, R.~P.\ 1994, \apj, 421, 753 

\bibitem[Stellingwerf(1978)]{ste78} Stellingwerf, R.~F.\ 
1978, \apj, 224, 953 

\bibitem[Thorstensen(1987)]{tho87} Thorstensen, J.~R.\ 1987, 
\apj, 312, 739 

\bibitem[Vieyro et al.(2012)]{vie12} Vieyro, F.~L., Sestayo, 
Y., Romero, G.~E., 
\& Paredes, J.~M.\ 2012, American Institute of Physics Conference Series, 1505, 410 

\bibitem[Wawrzyn et 
al.(2009)]{waw09} Wawrzyn, A.~C., Barman, T.~S., G{\"u}nther, H.~M., Hauschildt, P.~H., \& Exter, K.~M.\ 2009, \aap, 505, 227 

\bibitem[White \& Marshall(1983)]{whi83} 
White, N.~E., \& Marshall, F.~E.\ 1983, \iaucirc, 3806, 2 

\bibitem[Wijnands et al.(2002)]{wij02} Wijnands, R., Miller, 
J.~M., \& van der Klis, M.\ 2002, \mnras, 331, 60 

\bibitem[Yaqoob et al.(1993)]{yaq93} Yaqoob, T., Ebisawa, K., 
\& Mitsuda, K.\ 1993, \mnras, 264, 411 

\end{thebibliography}
\end{document}